\documentclass{article}

\usepackage[preprint, nonatbib]{neurips_2019}

\usepackage[utf8]{inputenc} 
\usepackage[T1]{fontenc}    
\usepackage{hyperref}       
\usepackage{url}            
\usepackage{booktabs}       
\usepackage{amsfonts}       
\usepackage{amsmath, amssymb}
\usepackage{nicefrac}       
\usepackage{microtype}      
\usepackage{graphicx}
\usepackage[labelformat=empty]{caption}
\usepackage[
citestyle=numeric
backend=biber,
style=numeric,
]{biblatex}
\title{Graph Nets for Partial Charge Prediction}

\author{%
  Yuanqing Wang\\
  Memorial Sloan Kettering Cancer Center\\
  New York, N.Y. 10065 USA\\
  \texttt{yuanqing.wang@choderalab.org} \\
  \And
  Josh Fass \\
  Memorial Sloan Kettering Cancer Center\\
  New York, N.Y. 10065 USA\\
  \texttt{josh.fass@choderalab.org} \\
  \And
  Chaya D. Stern \\
  Memorial Sloan Kettering Cancer Center\\
  New York, N.Y. 10065 USA\\
  \texttt{chaya.stern@choderalab.org} \\
  \And
  Kun Luo\\
  Uli Statistical Learning\\
  Shenzhen, Guangdong 518057, P.R. China\\
  \texttt{k@uli.ai}\\
  \AND
  John D. Chodera\thanks{Corresponding Author}\\
  Memorial Sloan Kettering Cancer Center\\
  New York, N.Y. 10065 \\
  \texttt{john.chodera@choderalab.org} \\
}

\DeclareUnicodeCharacter{2009}{ }
\DeclareUnicodeCharacter{0301}{\'{e}}

\begin{document}
\maketitle
\begin{abstract}
Atomic partial charges are crucial parameters for Molecular Dynamics (MD) simulations, molecular mechanics calculations, and virtual screening, as they determine the electrostatic contributions to interaction energies. 
Current methods for calculating partial charges, however, are either slow and scale poorly with molecular size (quantum chemical methods) or unreliable (empirical methods). 
Here, we present a new charge derivation method based on Graph Nets---a set of \textit{update} and \textit{aggregate} functions that operate on molecular topologies and propagate information thereon---that could approximate charges derived from Density Functional Theory (DFT) calculations with high accuracy and an over 500-fold speed up. 
\end{abstract}
\section{Introduction}
Molecular machine learning, with its significance in drug and materials design gradually being realized by the community, has been maturing rapidly. Recently, the focus has shifted from embedding molecules in a fixed-dimensional feature space to fit architectures borrowed from neighboring fields toward developing representations that respect the combinatorial, discrete nature of molecular structures. Among the newly developed algorithms are a plethora of graph-based methods~\cite{DBLP:journals/corr/GilmerSRVD17, kearnes2016molecular, DBLP:journals/corr/abs-1803-04465}, which represent molecules as graphs, atoms as nodes, and bonds as edges. Generally speaking, such representation makes it possible for the operations on modelled molecules to preserve the permutation invariance and equivariance with regard to chemically equivalent atoms. 

Despite the promising results achieved by these architectures in predicting per-molecule attributes (hydration free energies, binding affinities, lipohilicities, etc.), little attention has been paid to per-atom and per-bond regression tasks. This is drastically different from the general applications of graph learning, where per-node and per-edge problems are far more common. The prediction of atomic partial charges, we believe, could serve as an interesting pivotal task: 
As commercially available compound libraries now exceed $10^9$ molecules~\cite{hoffmann2019next}, there is a significant need for fast methods for determining high-quality partial charges for small organic molecules for virtual screening. 

In addition, next-generation force fields that consistently treat small molecules and biomolecules would greatly benefit from a fully consistent approach to determining charges for biopolymers (both naturally occurring and covalently modified) that exceed sizes practical for current quantum chemical calculations without requiring fragmentation and capping. This task may be particularly suited to machine learning methods due to the relative ease of generating an arbitrarily large number of partial charge training examples derived from high-level quantum chemistry. 

Our model was built on top of the full-block Graph Net structure~\cite{DBLP:journals/corr/abs-1806-01261}, where edges and the global node are attributed and connected to the nodes. Maximal generality and flexibility are provided by this framework so that, with certain choices of update and aggregate functions, the network could specialize into either a Message Passing Neural Net\cite{DBLP:journals/corr/GilmerSRVD17} or Graph Convolutional Network\cite{kearnes2016molecular}. We made contributions by designing and tuning its component functions while preserving the permutation equivariance end-to-end, as well as enforcing a \textit{net charge constraint}---the sum of the atomic partial charges equal the known total charge of the molecule---as a tractable quadratic program. We also optimized the batching pipeline for molecular graphs. This architecture is able to estimate charges derived from DFT calculations with an RMSE of $0.0223$~e on a subset of molecules from ChEMBL~\cite{10.1093/nar/gkr777} curated by Bleiziffer et al.~\cite{doi:10.1021/acs.jcim.7b00663}. Also, within the dataset, the prediction accuracy does not decrease as the size of the system increases. We therefore believe that this method, with results in agreement with quantum chemical calculations and computation time similar to empirical methods, has the potential to replace existing charge derivation schemes widely used in MD calculations and virtual screening approaches. Compared to the Random Forest model proposed by Bleiziffer et al.~\cite{doi:10.1021/acs.jcim.7b00663}, our model does not depend on rule-based fingerprints and is differentiable everywhere, and could therefore be optimized concurrently with forcefield parameters.

This work is part of a project that uses graph representations in molecular learning, termed GIMLET (Graph Inference on MoLEcular Topology). The infrastructure was implemented in Python~3.6 with no dependencies other than TensorFlow~2.0~\cite{tensorflow2015-whitepaper}, in order to ensure the pipelines can be integrated into TensorFlow computation graphs, and executed in a parallel and distributed fashion. The code is open source (\url{https://github.com/choderalab/gimlet}) and available under the MIT License.
\section{Architecture} \subsection{Molecules as graphs}
A \textit{graph} can be defined as a tuple of three sets:\begin{equation}
\mathcal{G} = \{ \mathcal{V, E, U}\}
\end{equation}where $\mathcal{V}$ is the set of the vertices (or nodes) (atoms), $\mathcal{E}$ the set of edges (bonds), and $\mathcal{U} = \{ \mathbf{u}\}$ the universal attribute. We model the molecules as node-, edge-, and globally attributed, undirected graphs, whose nodes are atoms and edges are bonds. In this notation, the connectivity is included in $\mathcal{E}$. In our implementation though, a vector $\mathbf{a}$ denoting a sequence of atomic numbers and a upper-triangular adjacency matrix $A$ is sufficient to describe the topology of the graph. \subsection{Graph Nets}
There are three stages in both the training and inference process: initialization, propagation, and readout, which, in our setting, are governed by sets of learnable functions.
\textit{Initialization stage} is the phase in which the data is abstracted as graph entities. Driven by the idea that the chemical environment of atoms could be realized in rounds of message passing, we featurize the atoms simply using one-hot encodings of the element identity, which is then fed into a feedforward neural network to form the vertex attributes at $t=0$, \begin{equation}
\mathbf{v}^{(-1)}_{ij} = \begin{cases}
1, &\text{atom } i \text{ is  element } j;\\
0, &\text{elsewhere,}\\
\end{cases}, \mathbf{v}^{(0)} = \mathrm{NN}_{v_0}(\mathbf{v}^{(-1)}) \in \mathbb{R}^{N_\text{atoms} \times d_v},
\end{equation}where $i \in \{1, 2, ..., N_\text{atoms}\}$ and $j \in \{1, 2, ..., N_\text{elements}\}$, and $d_v$ is the fixed dimension of node attributes. Edge attributes are initialized as the output of a feedforward network which takes the bond order (averaged over resonance structures) as input; global attributes are initialized as zeros as placeholder. \begin{equation}
\mathbf{e}^{(0)} = \mathrm{NN}_{e_0}(\mathrm{BO})\in \mathbb{R}^{N_\text{bonds} \times d_e},
\mathbf{u}^{(0)} = \mathbf{0}^{1 \times d_u},
\end{equation}where $d_e$ is the hidden dimension of edges and $d_u$ is the hidden dimension of global attributes.
In \textit{propagation stage}, the framework we adopted follows a formalism by Battaglia et al,\cite{DBLP:journals/corr/abs-1806-01261} where, in each round of message passing, the attributes of nodes, edges, and the graph as a whole, $\mathbf{v}, \mathbf{e}, \text{and } \mathbf{u}$ are updated by trainable functions in the following order:
    \begin{align}
    \mathbf{e}_k^{(t+1)} &= \phi^e(\mathbf{e}_k^{(t)}, \sum_{i \in \mathcal{N}^e_k}\mathbf{v}_i, \mathbf{u}^{(t)}), &\text{(edge update)}\\
    \bar{\mathbf{e}}_i^{(t+1)} &=\rho^{e\rightarrow v}(E_i^{(t+1)}), &\text{(edge to node aggregate)}\\
    \mathbf{v}_i^{(t+1)} &= \phi^v(\bar{\mathbf{e}}_i^{(t+1)}, \mathbf{v}_i^{(t)}, \mathbf{u}^{(t)}), &\text{(node update)}\\
    \bar{\mathbf{e}}^{(t+1)} &= \rho^{e \rightarrow u}(E^{(t+1)}), &\text{(edge to global aggregate)}\\
    \bar{\mathbf{v}}^{(t+1)} &= \rho^{v \rightarrow u}(V^{(t)}), &\text{(node to global aggregate)}\\
    \mathbf{u}^{(t+1)} &= \phi^u(\bar{\mathbf{e}}^{(t+1)}, \bar{\mathbf{v}}^{(t+1)}, \mathbf{u}^{(t)}), &\text{(global update)}
    \end{align}where $E_i=\{ \mathbf{e}_k, k\in \mathcal{N}_i^v\}$ is the set of attributes of edges connected to a specific node, $E_i = \{ e_k, k \in 1, 2, ..., N^e\}$ is the set of attributes of all edges, $V$ is the set of attributes of all nodes, and $\mathcal{N}^v$ and $\mathcal{N}^e$ denote the set of indices of entities connected to a certain node or a certain edge, respectively. $\phi^e$, $\phi^v$, and $\phi^u$ are update functions that take the \textit{environment} of the an entity as input and update the attribute of the entity, which could be stateful (Recurrent Neural Networks) or not; $\rho^{e \rightarrow v}$, $\rho^{e \rightarrow u}$, and $\rho^{v \rightarrow u}$ are aggregate functions that aggregate the attributes of multiple entities into an \textit{aggregated} attribute which shares the same dimension with each entity. Although in this work, the definition of edges is limited to that connect exactly two nodes (bonds connecting two atoms), we could expand the notion of edges to include \textit{hyperedges}, to connect more than two nodes (angles and torsions). 
    
Finally, after a designated number of rounds of propagation (message passing), in the \textit{readout stage}, $t=T$, a readout function $f^r$ that takes the entire trajectory as input summarizes the information and yields the final output of desired dimensionality,
\begin{equation}
\hat{y} = f^r(\{\{\mathbf{v^{(t)}}, \mathbf{e}^{(t)}, \mathbf{u}^{(t)}\}, t=1, 2,..., T\}).
\end{equation}\subsection{Graph Batching}
The number of nodes (atoms) in molecule graphs varies greatly and is usually much smaller than, say, the number of individuals in a social graph. For efficient backpropagation, especially on GPUs, molecule graphs need to be \textit{combined} into larger ones, rather than \textit{partitioned} or \textit{padded} to the same size. This could be achieved by concatenating the attribute vectors of graphs and merging their adjacency matrices of graphs as \begin{equation}
\widetilde{A}_{kl}=\begin{cases}
(\{A\}_j)_{k-\sum\limits_{m < j}|\mathcal{V}_m|, l-\sum\limits_{m < j}|\mathcal{V}_m|},
\text{where} \sum\limits_{m < j} |\mathcal{V}_m| \leq k, l < \sum\limits_{m < j+1} |\mathcal{V}_m|;\\
0, \text{ elsewhere}.
\end{cases}
\end{equation}After choosing an appropriate batch size, which is the first dimension of $\widetilde{\mathcal{V}}$ and $\widetilde{\mathcal{A}}$, we repeat this process until another addition of small graph into the batch would result in $\sum\limits_i |\mathcal{V}_i|$ greater than the batch size, upon which the adjacency and the concatenated attributes will be padded to the batch size and another batch will be initialized.
\subsection{Determination of atomic partial charges respecting a net charge constraint}
One of the challenges in predicting atomic partial charges is to satisfy the constraint that their sum should equal to the total charge of the molecule:
\begin{equation}\label{constraint}
\sum_i \hat{q}_i = \sum_i q_i = Q,
\end{equation}where $Q$ is the total (net) charge of the molecule, which could be positive, negative, or zero. Naively, we could either not explicitly encode this constraint and let the model "learn" it, or, as in Bleiziffer et al.~\cite{doi:10.1021/acs.jcim.7b00663}, redistribute charge necessary to cancel any "excess charge" evenly to all atoms. Experimentally, none of these methods achieved satisfactory results when used with our model (with no constraint, RMSE is around 0.280 e.) We instead adopted a trick proposed by Gilson et al.~\cite{doi:10.1021/ci034148o} and use our model to instead predict the first- and second-order derivatives of the potential energy $E$ w.r.t.\ the atomic partial charge, which happens to correspond to the \textit{electronegativity} $e_i$ and \textit{hardness} $s_i$ of the atom in its chemical environment.\begin{equation}
        e_i \equiv \frac{\partial E}{\partial q_i},
        s_i  \equiv \frac{\partial E^2}{\partial^2 q_i}.
    \end{equation} 
This problem could thus be formulated as follows: we use the graph net to make a prediction of the electronegativity and hardness, $\{\hat{e}_i, \hat{s}_i\}$, and the partial charges could be yielded by minimizing the second-order Taylor expansion of the potential energy contributed by atomic charges:  \begin{equation}\label{q}
        \{ \hat{q}_i\} = \underset{q_i}{\mathrm{argmin}}\sum_i \hat{e}_i q_i + \frac{1}{2}\hat{s}_i \hat{q}_i^2,
    \end{equation}subject to \ref{constraint}. Fortunately, using Lagrange multipliers, the solution to \ref{q} could be given analytically by:
    \begin{equation}
        \hat{q}_i = - e_i s_i^{-1} + s_i^{-1}\frac{Q + \sum_i e_i s_i^{-1}}{\sum_j s_j ^{-1}},
    \end{equation}whose Jacobian and Hessian are trivially easy to calculate. As a result, the prediction of $\{\hat{e}_i, \hat{s}_i\}$ could be optimized end-to-end using backpropagation.
\section{Results and Discussion}
\begin{figure}[htbp]
\begin{minipage}[tb]{0.4\linewidth}
    \centering
    \includegraphics[width=\textwidth]{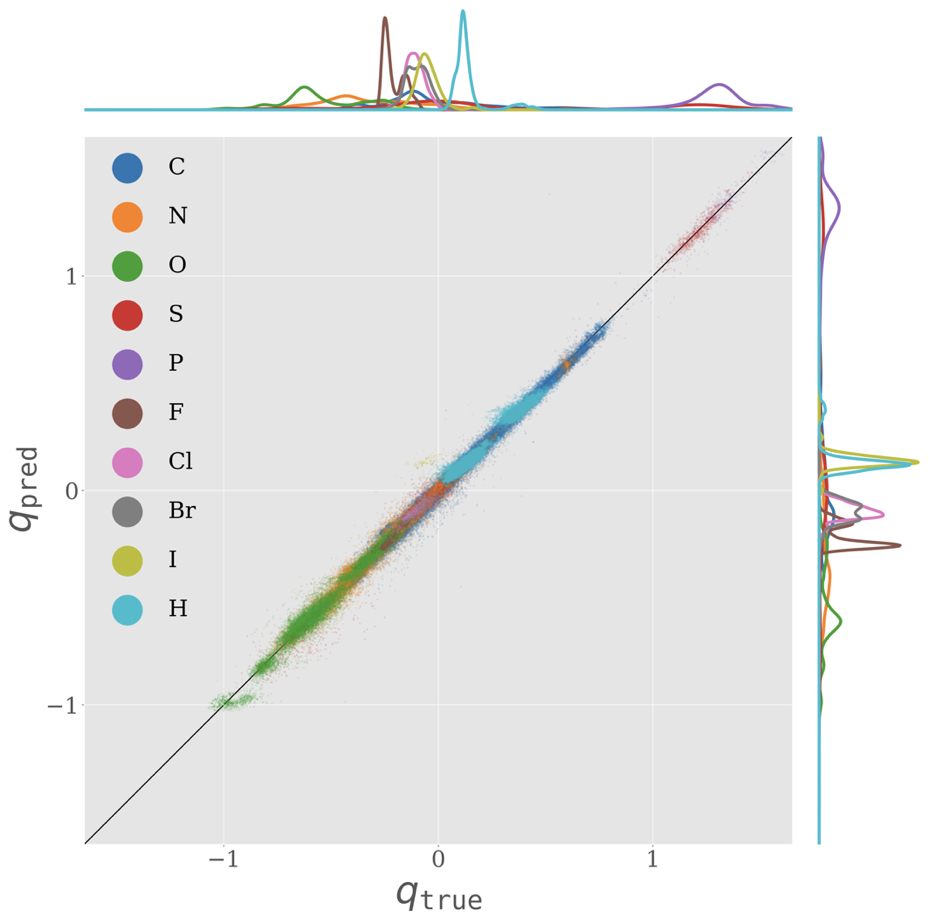}
\end{minipage}
\begin{minipage}[tb]{0.4\linewidth}
    \small
    \centering
    \begin{tabular}[\textwidth]{c c c c c}
    \hline
    Element & $R^2$ & RMSE(e)& \# Samples \\
    \hline
    C & $ 0.9932_{0.9930}^{0.9933} $ & $ 0.0217_{0.0215}^{0.0219} $ & 116864\\
    N & $ 0.9797_{0.9789}^{0.9805} $ & $ 0.0370_{0.0364}^{0.0376} $ & 19490\\
    O & $ 0.9713_{0.9700}^{0.9725} $ & $ 0.0342_{0.0336}^{0.0348} $ & 21503\\
    S & $ 0.9935_{0.9928}^{0.9942} $ & $ 0.0524_{0.0496}^{0.0551} $ & 2955\\
    P & $ 0.8582_{0.7265}^{0.9943} $ & $ 0.0669_{0.0339}^{0.0950} $ & 341\\
    F & $ 0.9517_{0.9458}^{0.9577} $ & $ 0.0132_{0.0126}^{0.0138} $ & 1967\\
    Cl & $ 0.7781_{0.7516}^{0.8049} $ & $ 0.0253_{0.0236}^{0.0270} $ & 1215\\
    Br & $ 0.8166_{0.7878}^{0.8458} $ & $ 0.0233_{0.0214}^{0.0252} $ & 572\\
    I & $ 0.2819_{-0.0178}^{0.6376} $ & $ 0.1948_{0.1874}^{0.2017} $ & 105\\
    H & $ 0.9744_{0.9739}^{0.9750} $ & $ 0.0144_{0.0142}^{0.0145} $ & 134799\\
    \hline
    Overall & $ 0.9936_{0.9935}^{0.9937} $ & $ 0.0223_{0.0221}^{0.0225} $ & 299811 \\
    \hline
    \end{tabular}
\end{minipage}
\caption{
\textbf{Figure 1 (Left): Predicted vs true partial charge of atoms in held-out test set color-coded by element types.}
A kernel density estimate of the distribution of charges for each element are plotted on the axes. \\
\textbf{Table 1 (Right): $\mathbf{R^2}$ and RMSE of the prediction and number of data points in held-out test set.} 
The 95\% confidence interval is also annotated.}
\end{figure}
We tested our model on a dataset consisting of 350 259 molecules in ChEMBL database, selected by Bleiziffer et al.~\cite{doi:10.1021/acs.jcim.7b00663} The reference charges are also calculated by Bleiziffer et al.~\cite{doi:10.1021/acs.jcim.7b00663} using DFT with dielectric permittivity $\epsilon=4$. We randomly split the training and test set with 80:20 ratio. Random search on a limited hyperparameter space was conducted for hyperparameter tuning, with the hyperparameter set with highest 5-fold cross validation results chosen. On the test set, the error between the true and predicted value, RMSE~$\approx0.02$~e, is roughly within the difference between DFT and AM1-BCC calculations, whereas it takes around 0.03~seconds to calculate the charges for a single molecule, which is approximately 500 times faster than AM1-BCC methods. We therefore argue that such method has the potential to replace AM1-BCC in calculating the charges for small molecules for MD simulation. Moreover, within the dataset (where the largest molecule has 63 atoms), we observed no positive correlation between the prediction error and the number of atoms in the molecule, indicating potential scalability of this model.

One potential problem of training our model on this dataset is that, charges yielded from DFT calculations are also dependent on the conformation of the molecule, although the conformation used in the calculation is thought to be of lowest energy, thus is deterministic with regard to the topology. Currently, we are using AM1-BCC ELF10 methods to calculate partial charges that could be seen as independent of conformation. Apart from charges, such calculations could also yield per-bond (per-edge), and per-molecule (per-graph) features, namely partial bond orders such as Wiberg bond order and formation energy. We plan to continue this study by carrying out experiments to predict these features independently, jointly as multitask learning, and predicting some with others given. 

\section{Acknowledgments}

JDC acknowledges support from the Sloan Kettering Institute and NIH grant P30 CA008748 from the National Cancer Institute.
JDC acknowledges support from the National Institute for General Medical Sciences of the National Institutes of Health under award number R01GM121505.
JDC is grateful to OpenEye Scientific for providing a free academic software license for use in this work.
JF acknowledges support from the National Science Foundation under award number NSF CHE-1738979.
CDS was supported by a fellowship from The Molecular Sciences Software Institute under NSF grant ACI-1547580.

\section{Disclosures}

JDC is a current member of the Scientific Advisory Board of OpenEye Scientific Software.
YW and KL are co-founders and equity holders of Uli Statistical Learning.
The Chodera laboratory receives or has received funding from multiple sources, including the National Institutes of Health, the National Science Foundation, the Parker Institute for Cancer Immunotherapy, Relay Therapeutics, Entasis Therapeutics, Silicon Therapeutics, EMD Serono (Merck KGaA), AstraZeneca, Vir Biotechnology, XtalPi, the Molecular Sciences Software Institute, the Starr Cancer Consortium, the Open Force Field Consortium, Cycle for Survival, a Louis V. Gerstner Young Investigator Award, and the Sloan Kettering Institute. 
A complete funding history for the Chodera lab can be found at \url{http://choderalab.org/funding}. 

\section{Disclaimers}

The content is solely the responsibility of the authors and does not necessarily represent the official views of the National Institutes of Health.

\printbibliography

@article{hoffmann2019next,
  title={The next level in chemical space navigation: going far beyond enumerable compound libraries},
  author={Hoffmann, Torsten and Gastreich, Marcus},
  journal={Drug discovery today},
  year={2019},
  publisher={Elsevier}
}

@article{DBLP:journals/corr/GilmerSRVD17,
  author    = {Justin Gilmer and
               Samuel S. Schoenholz and
               Patrick F. Riley and
               Oriol Vinyals and
               George E. Dahl},
  title     = {Neural Message Passing for Quantum Chemistry},
  journal   = {CoRR},
  volume    = {abs/1704.01212},
  year      = {2017},
  url       = {http://arxiv.org/abs/1704.01212},
  archivePrefix = {arXiv},
  eprint    = {1704.01212},
  bibsource = {dblp computer science bibliography, https://dblp.org}
}

@article{kearnes2016molecular,
  title={Molecular graph convolutions: moving beyond fingerprints},
  author={Kearnes, Steven and McCloskey, Kevin and Berndl, Marc and Pande, Vijay and Riley, Patrick},
  journal={Journal of computer-aided molecular design},
  volume={30},
  number={8},
  pages={595--608},
  year={2016},
  publisher={Springer}
}

@article{DBLP:journals/corr/abs-1803-04465,
  author    = {Evan N. Feinberg and
               Debnil Sur and
               Brooke E. Husic and
               Doris Mai and
               Yang Li and
               Jianyi Yang and
               Bharath Ramsundar and
               Vijay S. Pande},
  title     = {Spatial Graph Convolutions for Drug Discovery},
  journal   = {CoRR},
  volume    = {abs/1803.04465},
  year      = {2018},
  url       = {http://arxiv.org/abs/1803.04465},
  archivePrefix = {arXiv},
  eprint    = {1803.04465},
  bibsource = {dblp computer science bibliography, https://dblp.org}
}

@article{DBLP:journals/corr/abs-1806-01261,
  author    = {Peter W. Battaglia and
               Jessica B. Hamrick and
               Victor Bapst and
               Alvaro Sanchez{-}Gonzalez and
               Vin{\'{\i}}cius Flores Zambaldi and
               Mateusz Malinowski and
               Andrea Tacchetti and
               David Raposo and
               Adam Santoro and
               Ryan Faulkner and
               {\c{C}}aglar G{\"{u}}l{\c{c}}ehre and
               H. Francis Song and
               Andrew J. Ballard and
               Justin Gilmer and
               George E. Dahl and
               Ashish Vaswani and
               Kelsey R. Allen and
               Charles Nash and
               Victoria Langston and
               Chris Dyer and
               Nicolas Heess and
               Daan Wierstra and
               Pushmeet Kohli and
               Matthew Botvinick and
               Oriol Vinyals and
               Yujia Li and
               Razvan Pascanu},
  title     = {Relational inductive biases, deep learning, and graph networks},
  journal   = {CoRR},
  volume    = {abs/1806.01261},
  year      = {2018},
  url       = {http://arxiv.org/abs/1806.01261},
  archivePrefix = {arXiv},
  eprint    = {1806.01261},
  bibsource = {dblp computer science bibliography, https://dblp.org}
}

@article{10.1093/nar/gkr777,
    author = {Gaulton, Anna and Bellis, Louisa J. and Bento, A. Patricia and Chambers, Jon and Davies, Mark and Hersey, Anne and Light, Yvonne and McGlinchey, Shaun and Michalovich, David and Al-Lazikani, Bissan and Overington, John P.},
    title = "{ChEMBL: a large-scale bioactivity database for drug discovery}",
    journal = {Nucleic Acids Research},
    volume = {40},
    number = {D1},
    pages = {D1100-D1107},
    year = {2011},
    month = {09},
    issn = {0305-1048},
    doi = {10.1093/nar/gkr777},
    url = {https://doi.org/10.1093/nar/gkr777},
    eprint = {http://oup.prod.sis.lan/nar/article-pdf/40/D1/D1100/16955876/gkr777.pdf},
}

@article{doi:10.1021/acs.jcim.7b00663,
author = {Bleiziffer, Patrick and Schaller, Kay and Riniker, Sereina},
title = {Machine Learning of Partial Charges Derived from High-Quality Quantum-Mechanical Calculations},
journal = {Journal of Chemical Information and Modeling},
volume = {58},
number = {3},
pages = {579-590},
year = {2018},
doi = {10.1021/acs.jcim.7b00663},
    note ={PMID: 29461814},

URL = { 
        https://doi.org/10.1021/acs.jcim.7b00663
    
},
eprint = { 
        https://doi.org/10.1021/acs.jcim.7b00663
    
}

}

@article{doi:10.1021/ci034148o,
author = {Gilson, Michael K. and Gilson, Hillary S. R. and Potter, Michael J.},
title = {Fast Assignment of Accurate Partial Atomic Charges:  An Electronegativity Equalization Method that Accounts for Alternate Resonance Forms},
journal = {Journal of Chemical Information and Computer Sciences},
volume = {43},
number = {6},
pages = {1982-1997},
year = {2003},
doi = {10.1021/ci034148o},
    note ={PMID: 14632449},

URL = { 
        https://doi.org/10.1021/ci034148o
    
},
eprint = { 
        https://doi.org/10.1021/ci034148o
    
}

}

@misc{tensorflow2015-whitepaper,
title={ {TensorFlow}: Large-Scale Machine Learning on Heterogeneous Systems},
url={https://www.tensorflow.org/},
note={Software available from tensorflow.org},
author={
    Mart\'{\i}n~Abadi and
    Ashish~Agarwal and
    Paul~Barham and
    Eugene~Brevdo and
    Zhifeng~Chen and
    Craig~Citro and
    Greg~S.~Corrado and
    Andy~Davis and
    Jeffrey~Dean and
    Matthieu~Devin and
    Sanjay~Ghemawat and
    Ian~Goodfellow and
    Andrew~Harp and
    Geoffrey~Irving and
    Michael~Isard and
    Yangqing Jia and
    Rafal~Jozefowicz and
    Lukasz~Kaiser and
    Manjunath~Kudlur and
    Josh~Levenberg and
    Dandelion~Man\'{e} and
    Rajat~Monga and
    Sherry~Moore and
    Derek~Murray and
    Chris~Olah and
    Mike~Schuster and
    Jonathon~Shlens and
    Benoit~Steiner and
    Ilya~Sutskever and
    Kunal~Talwar and
    Paul~Tucker and
    Vincent~Vanhoucke and
    Vijay~Vasudevan and
    Fernanda~Vi\'{e}gas and
    Oriol~Vinyals and
    Pete~Warden and
    Martin~Wattenberg and
    Martin~Wicke and
    Yuan~Yu and
    Xiaoqiang~Zheng},
  year={2015},
}
\end{document}